\documentstyle[aps,prc,prl,multicol]{revtex}

\newcommand{\AmS}{{\protect\the\textfont2
  A\kern-.1667em\lower.5ex\hbox{M}\kern-.125emS}}

\begin{document}
\draft
\title{First Measurement of the 
$^{3}He(^{3}He,2p)^{4}He$ Cross Section down to the Lower Edge of
the Solar Gamow Peak}

\author{
THE LUNA COLLABORATION \\
R. Bonetti$^1$,
C. Broggini$^{2,*}$,
L. Campajola$^3$,
P. Corvisiero$^4$,
A. D'Alessandro$^5$,
M. Dessalvi$^4$,
A. D'Onofrio$^6$,
A. Fubini$^7$,
G. Gervino$^8$,
L. Gialanella$^9$
U. Greife$^9$,
A. Guglielmetti$^1$,
C. Gustavino$^5$,
G. Imbriani$^3$,
M. Junker$^5$,
P. Prati$^4$,
V. Roca$^3$,
C. Rolfs$^9$,
M. Romano$^3$,
F. Schuemann$^9$,
F. Strieder$^9$,
F. Terrasi$^3$,
H.P. Trautvetter$^9$,
S. Zavatarelli$^4$}
\address{
$^1 $ Universit\'{a} di Milano,  Dipartimento di Fisica and INFN, Milano\\
$^2 $ INFN, Padova                  \\
$^3 $ Universit\'{a} di Napoli "Federico II",  Dipartimento 
di Scienze Fisiche and INFN, Napoli\\
$^4 $ Universit\'{a} di Genova, Dipartimento di Fisica and INFN, Genova\\
$^5 $ Laboratori Nazionali del Gran Sasso, Assergi\\
$^6 $ Seconda Universit\'{a} di Napoli, Dipartimento di Scienze Ambientali, 
Caserta and INFN, Napoli\\
$^7 $ ENEA, Frascati and INFN, Torino\\
$^8 $ Universit\'{a} di Torino, Dipartimento di Fisica Sperimentale and INFN, 
Torino\\
$^9 $ Institut f\"ur Experimentalphysik III, Ruhr--Universit\"at Bochum}
\date{\today}
\maketitle
\begin{abstract}
We give the LUNA results on the 
$\sigma(E)$
cross section measurement
of a key reaction of the proton-proton chain strongly affecting the
calculated neutrino luminosity from the Sun:
$^{3}He(^{3}He,2p)^{4}He$.
Due to the cosmic ray suppression provided by the Gran Sasso underground
laboratory it has been possible to measure 
$\sigma(E)$ down to the lower edge of 
the solar Gamow peak, i.e. as low as  
16.5 $keV$ centre of mass energy.
The data clearly show
the cross section
increase due 
to the electron 
screening effect but they do not exhibit any
evidence for a narrow resonance suggested to explain the observed 
solar neutrino flux.
\end{abstract}
\pacs{PACS numbers: 26.65.+t, 25.90.+k}
\begin{multicols}{2}
The nuclear reactions which generate the energy of stars and,
in doing so,  synthesize 
elements occur inside stars  
at energies within the Gamow peak: $E_0 \pm \delta E_0$.
In this region, which is
far below the Coulomb energy 
$E_{\rm c}$
(approximately $E_{\rm 0}/E_{\rm c} = 0.01$), the reaction cross section 
$\sigma(E)$ drops nearly exponentially with decreasing energy $E$ 
\cite{rol88}:
\begin{equation}
\sigma(E)= \frac{S(E)}{E}\,{exp(-2\,\pi\, \eta)}, \label{yielddef1}
\end{equation}
where $S(E)$ is the astrophysical factor and
$\eta$ is the Sommerfeld parameter, given by
$2\, \pi\, \eta=31.29\, Z_1\, Z_2(\mu/E)^{1/2}$.
$Z_1$ and $Z_2$  are  the 
nuclear charges of the interacting particles  in  the  entrance channel,
$\mu$ is the reduced mass (in units of amu),  and  $E$  is  the  center 
of mass
energy  (in   units  of  keV). 

The extremely low value of $\sigma(E)$ within the Gamow peak has always
prevented
its measurement in a laboratory at the Earth surface. As a matter of fact, 
the signal to background ratio would be too small because of the cosmic ray 
interactions.
Instead, the observed 
energy dependence of $\sigma(E)$ at high energies is extrapolated to 
the low energy region, leading to substantial uncertainties. In particular, a 
possible resonance in the unmeasured region is not accounted for by the 
extrapolation, but it could completely dominate the reaction rate at the 
Gamow peak. 

In addition another effect can be studied at low energies:
the electron 
screening. The beam and the target used in an experiment are usually 
ions and neutral atoms, respectively. The electron clouds surrounding 
the interacting nuclei act as a screening potential, thus reducing the 
height of the Coulomb barrier and 
leading to a higher cross section,
$\sigma_{\rm s}(E)$, than would  be  the case for bare nuclei, 
$\sigma_{\rm b}(E)$,  with an exponential  enhancement  factor 
$\cite{rol88}$:

\begin{equation}
f_{\rm lab}(E) = \sigma_{\rm s}(E)/
\sigma_{\rm b}(E) \simeq \exp(\pi\eta\, U_{\rm e}/E), \label{yielddef2}
\end{equation}  
where $U_{\rm e}$ 
is the electron-screening potential energy.
It should be pointed out that 
the screening effect has to be measured and taken into account to 
derive the bare nuclei cross section, which is the 
input data to the models of stellar nucleosynthesis.

Therefore both the search for narrow resonances and the study of 
electron screening demand for
the direct measurement of the nucleosynthesis cross sections
in the low energy region (few tens of $keV$).
In order to start exploring this new and fascinating domain of 
nuclear astrophysics we installed an accelerator 
facility deeply underground 
where the cosmic rays, which are the limiting background in all the 
existing experiments, are strongly suppressed.

LUNA \cite{arp91}
(Laboratory for Underground Nuclear Astrophysics) 
is located in
a dedicated room 
of the Laboratori Nazionali del Gran Sasso
(LNGS), separated from other experiments by at least 60 $m$ of 
rock. The mountain provides a natural shielding equivalent to
at least 3800 meters of water which reduces the muon and neutron fluxes 
by a 
factor $10^{6}$ and $10^{3}$, respectively. The $\gamma$ ray flux is like
the surface one,
but a detector can be more effectively shielded underground 
due to the suppression of the cosmic ray induced background.

Technical details of the LUNA set-up have already been reported $\cite{gre94}$. 
Briefly, the
50 kV accelerator facility 
consists of a duoplasmatron ion source,
an extraction/acceleration system, a double-focusing 90$^{\rm o}$ analysing
magnet, a windowless gas-target system and a beam
calorimeter.
Its outstanding features are the following: very small beam energy spread
(the source spread is less than 20 $eV$,  
acceleration voltage known with an accuracy of better than $10^{-4}$),
and
high beam current even at low energy (about 300 $\mu$A
measurable with a 3$\%$ accuracy).

Since the beginning 
LUNA has been focused 
on the $^{3}He(^{3}He,2p)^{4}He$ 
cross section measurement within the solar 
Gamow peak (15-27 $keV$).
This reaction plays a big role in the proton-proton chain,
strongly affecting the calculated solar neutrino luminosity. 
As a matter of fact a  resonance at the thermal energy of the Sun 
has been suggested 
long time ago  
\cite{fow72} \cite{fet72} to explain the observed 
solar neutrino flux, which is a factor between 2 and 3 lower than  
expected. The possible resonant enhancement in the 
$^{3}He(^{3}He,2p)^{4}He$ 
cross section would decrease the relative contribution of the alternative 
reaction  
$^{3}He(\alpha,\gamma)^{7}Be$, which generates the branch responsible for
$^{7}Be$ and $^{8}B$
neutrino production in the Sun.

A resonance at an energy far below 100 $keV$ has also been discussed
\cite{gal94}
to explain the galactic abundance of  $^{3}He$. It is known \cite{ree73}
that big-bang 
nucleosynthesis alone generates enough $^{3}He$ to account for the 
observations. The $^{3}He$ production by stars is not required: the 
resonance in the  
$^{3}He(^{3}He,2p)^{4}He$ 
cross section could provide a mechanism through which the produced 
$^{3}He$ is destroyed inside stars. 

Before LUNA the 
$^{3}He(^{3}He,2p)^{4}He$ 
cross section measurements stopped at the centre of mass 
energy of 24.5 $keV$ 
($\sigma$=7$\pm$2$pb$)\cite{kra87}, just at the
upper edge of the thermal energy region of the 
Sun. In the underground experiment we measured 
for the first time down to 20.76 $keV$
$\cite{arp96,jun98}$ and then, with a different 
detector set-up, down to 16.5 $keV$.

The first detector set-up consisted of 
four ${\Delta}$E-E telescopes
placed around the beam axis.
Each telescope was made of a thin (140 $\mu$m) transmission surface barrier
silicon detector followed by a thick one (1 $mm$), both of
5x5 $cm^{2}$ area.
The signature of a 
$^{3}He(^{3}He,2p)^{4}He$ event  
was a proton signal within a fixed energy 
window in one and only one telescope. 

With this set-up we collected data 
until August 1996 down to the centre of mass energy of 20.76 $keV$,
with no evidence of the hypothetical 
resonance.
At the lowest energy  
we had a rate of 3 events/day, giving a cross section
$\sigma$=0.9$\pm$0.1$pb$.

We could not go further down because of the background reaction 
$^{3}He(d,p)^{4}He$, which at low energies has a cross section 
much larger than that of 
$^{3}He(^{3}He,2p)^{4}He$ 
(deuterium is contained in the $^{3}He^{+}$ beam as $HD^{+}$ molecule):
for instance, at 16.5 $keV$ centre of mass energy it is larger by eight orders
of magnitude.

The 14.9 $MeV$ protons from
$^{3}He(d,p)^{4}He$
have an energy 
larger than the protons from
$^{3}He(^{3}He,2p)^{4}He$ 
(a continuos spectrum up to 10.7 MeV), however
a few of them
can hit the detectors
near the edges of their active volumes  loosing  only  a fraction of their
energy and  thus giving the same signature as
the $^{3}He(^{3}He,2p)^{4}He$ events.

In order to suppress this background we had to use a new detector 
set-up, which consists of eight thick (1 $mm$) silicon detectors of 
5x5 $cm^{2}$ area 
placed around the beam. They form a 12 $cm$ long 
parallelepiped in the 
target chamber, at 5.3 $cm$ from its entrance.
Each detector is cooled down to -20 $^{\circ}C$,
to reduce the leakage current, and it is shielded by  
a 1 $\mu$m thick Mylar foil, 
a 1 $\mu$m aluminium foil 
and a 10 $\mu$m nickel cylinder in order to stop 
the produced $^{4}He$ nuclei,
the elastically scattered $^{3}He$ and the light 
induced by the beam.
 
Standard NIM electronics is reading out the detectors. The
signals are then  
handled   and   stored   using  a   CAMAC  multiparametric   system
which also stores information on the experimental
parameters
and on the count rate of the pulser used for
dead time and
electronic stability checks. 

Inside the chamber, which has a length of 41.9 $cm$, there is 
a constant $^{3}He$ gas pressure of
0.5 $mbar$
(measured 
to an accuracy of better than 1\%). In going through the gas, the beam
experiences a mean energy loss of about 3 $keV$ (1 $keV$  to the
middle of the detector set-up). This is taken into account   
by introducing an effective beam energy
$E_{\rm eff}$ corresponding to the mean value of the beam energy
distribution in the detector set-up, evaluated by 
Monte Carlo simulation for each different accelerating voltage.
Since at subcoulomb energies a precise knowledge of the effective beam energy
is crucial, 
all the Monte Carlo predictions 
have been thoroughly
tested by changing the target gas pressure and the detector position.

In the data analysis we 
want to select those events where two protons are detected. As a
matter of fact this is the 
signature which unambiguosly identifies a
$^{3}He(^{3}He,2p)^{4}He$ fusion reaction, thus 
completely suppressing 
the background events due to $^{3}He(d,p)^{4}He$
where only one proton is emitted. It was not possible to ask for such a
signature in the old set-up because the detection efficiency would have been
too
low (less than 1$\%$).

Therefore selected events 
must fulfill the
following  conditions:
\begin{enumerate}
\item there is a coincidence, within 1 $\mu$s, between the 
signals of two silicon detectors;
this 
essentially eliminates the events due to the 
natural  radioactivity of the detectors themselves and of the surrounding 
materials;
\item each proton deposits more than 2 $MeV$ in the detector
and the sum of the two proton energies is within the constraints
given by the Q-value (12.86 $MeV$) of the reaction,
thus 
cutting away the electronic  noise;
\item  the coincidence occurs between two and only two detectors;
in this way events which
trigger more than two detectors are rejected in order 
to remove the residual
electronic noise and the muon induced showers.
\end{enumerate}

These requirements lead 
to an absolute detection efficiency of 5.3$\pm$ 0.2$\%$
as determined by the Monte Carlo program \cite{arp95}. 
No event has been detected fulfilling our selection criteria 
during a 23 day background run with a $^{4}He$ beam 
(297 $C$) on
a $^{4}He$ target (0.5 $mbar$). 

In Table I we give the new results
which conclude the LUNA measurement of 
the $^{3}He(^{3}He,2p)^{4}He$ cross section.
We point out that 
the cross section varies by more than two orders of magnitude in the 
measured energy range. At the lowest energy of 16.5 $keV$ it has 
the value of
0.02$\pm$0.02$pb$,
which corresponds to
a rate 
of about 2 events/month, rather low even for the 'silent' 
experiments of underground physics.

The astrophysical factor $S(E)$ is also shown 
in Fig. 1,
together with 
the values we obtained underground with the four telescope set-up:
there is an excellent agreement between the 
two different detector set-ups in the overlapping region 

The dominant error on the new data is the statistical one,
with the systematical error mainly arising from
the 10$\%$ uncertainty in the beam energy loss inside the target.

In Fig. 2
we plot two existing measurements 
\cite{kra87}\cite{dwa71} 
of the 
astrophysical factor $S(E)$ together with our underground 
and surface \cite{jun98} results.
By fitting the observed energy dependence of $S(E)$
 from 16.5 $keV$ to 1080 $keV$ with the expressions:
\begin{eqnarray}
S_{\rm b}(E) & = & S_{\rm b}(0) + S_{\rm b}'(0)E + 0.5\, S_{\rm b}''(0)E^2  \\
S_{\rm s}(E) & = & S_{\rm b}(E) \exp(\pi \eta U_{\rm e}/E),
\end{eqnarray}
we obtain the values of the parameters 
$S_{\rm b}(0)$, $S_{\rm b}'(0)$, $S_{\rm b}''(0)$ and $U_{\rm e}$ 
given in 
Table II
($S_{\rm b}$ 
and $S_{\rm s}$ are the astrophysical factors for bare and shielded nuclei,
respectively).

We observe that these
values are in excellent agreement with our previous results \cite{jun98}
and that the
screening potential (294$\pm$47 $eV$) is close to
the one from the adiabatic
limit (240 eV).

From our measurement it is concluded
that 
the $^{3}He(^{3}He,2p)^{4}He$ cross section
increases at the thermal energy of the Sun as  expected from
the electron 
screening effect but does not show any
evidence for a narrow resonance.
Consequently, the astrophysical solution of the solar neutrino problem based on
its existence is ruled out by our results. 

In conclusion, LUNA has provided the first 
cross section measurement
of a key reaction of the proton-proton chain
at the thermal energy of the Sun. In this way it has
also shown that, by going underground and by using 
the
typical techniques of low background physics, it is possible to 
measure 
nuclear cross sections down to the energy of the nucleosynthesis
inside stars.

We wish to thank the Director and the staff
of the Laboratori Nazionali del Gran Sasso for their support and 
hospitality.

\end{multicols}
\clearpage
\begin{figure}
\caption{The $^{3}He(^{3}He,2p)^{4}He$ astrophysical factor $S(E)$
measured underground with the LUNA old set-up
(four telescopes) and with the new one (eight thick silicon detectors).
The error bars correspond to one standard deviation.}
\label{luna1}
\end{figure}      
\begin{figure}
\caption{The $^{3}He(^{3}He,2p)^{4}He$ astrophysical factor $S(E)$
from two previous measurements and 
from LUNA (underground $+$ surface). The lines are the fit
to the astrophysical factors of bare and shielded nuclei.
The solar Gamow peak is shown in arbitrary units.}
\label{luna2}
\end{figure}      
\newpage
\begin{table}[h]
\caption{The LUNA results with the new detector set-up.}
\begin{tabular}{cccccc}
%
%
\multicolumn{1}{c}{Energy \tablenotemark[1]}  &
\multicolumn{1}{c}{Charge \tablenotemark[2]}  &
\multicolumn{1}{c}{Events }  &
\multicolumn{1}{c}{S(E) \tablenotemark[3]} &
\multicolumn{1}{c}{$\Delta$S$_{\rm stat}$ \tablenotemark[4]} &
\multicolumn{1}{c}{$\Delta$S$_{\rm sys}$ \tablenotemark[5]} \\
\multicolumn{1}{c}{  (keV) }    &
\multicolumn{1}{c}{   (C) }     &      
\multicolumn{1}{c}{}  &
\multicolumn{1}{c}{(MeV b)} &
\multicolumn{1}{c}{(MeV b)} &
\multicolumn{1}{c}{(MeV b)}\\
\hline
%
16.50 
&    349   &   1  &  7.70 &   7.70  &  0.49   \\
16.99
&    827   &   7  & 13.15 &   4.98  &  0.83   \\
  17.46
&    189   &   1  &  5.26 &   5.26  &  0.33    \\
  17.97
&    272   &   0  &  $<$14   &         &         \\
  18.46
&    337   &   7  &  7.86 &   2.97  &  0.47     \\
  18.98
&    387   &  13  &  8.25 &   2.29  &  0.48     \\
  19.46
&    242   &  12  &  7.67 &   2.22  &  0.44     \\
  19.93
&    190   &   9  &  5.10 &   1.70  &  0.29     \\
  21.43
&    365   &  53  &  4.72 &   0.65  &  0.26     \\
  23.37 
&    167   & 141  &  7.31 &   0.63  &  0.39     \\
  24.36
&    298   & 278  &  5.44 &   0.34  &  0.28     \\
%
\end{tabular}
\tablenotetext[1]
{Effective center of mass energy derived from the absolute energy of the
ion beam and  the Monte  Carlo
simulation (including the  energy loss of the beam inside the target
gas and the effects of the extended gas-target and detector geometries).}
\tablenotetext[2]
{Deduced from the beam calorimeter.}
\tablenotetext[3]
{The upper limit at 17.97 $keV$ energy is given at the 95 $\%$ 
confidence level.}
\tablenotetext[4]
{Statistical error (one standard deviation).}
\tablenotetext[5]
{Systematical error (one standard deviation).}
\end{table}
\begin{table}
\caption{The $S_{\rm b}(E)$ factors and the electron
screening potential energy $U_{\rm e}$ given by the fit to the data shown in
Fig. 2.}
\begin{tabular}{ccccc}
%
%
{$S_{\rm b}(0)$}  &
{$S_{\rm b}'(0)$}  &
{$S_{\rm b}''(0)$}  &
{$U_{\rm e}$} &
{$\chi^2$/$d.o.f.$}  \\
{(MeV b)}  &
{(b)}  &
{(b/MeV)}  &
{(eV)} &\\
\hline
 $5.32 \pm 0.08$ & $-3.7 \pm 0.6$ & $3.9 \pm 1.0$ & $294 \pm 47$  & 0.86 \\%
\end{tabular}
\end{table}
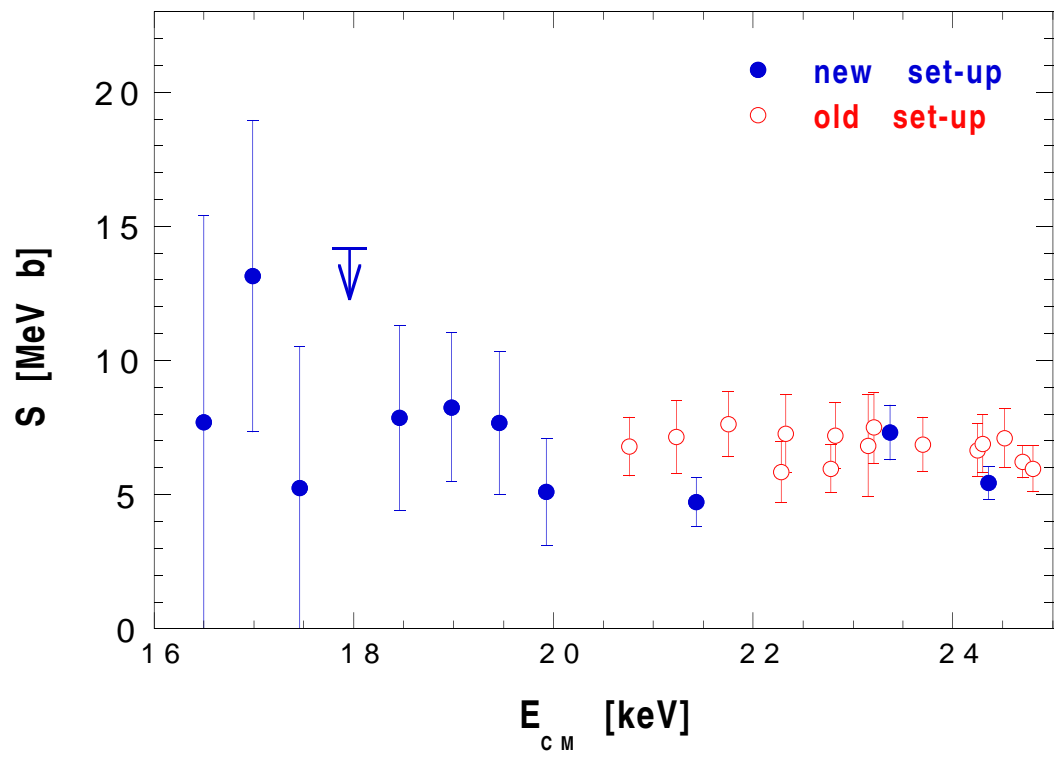
\end{document}